\documentclass{article}
\usepackage{frascatiphys,here,graphicx,subfigure}
\begin{document}
\title{ 
STUDY OF THE DECOHERENCE OF ENTANGLED KAONS
BY THE INTERACTION WITH THERMAL PHOTONS
}
\author{
Izabela Balwierz\\
{\em Jagiellonian University, Institute of Physics, Krakow, Poland} \\
Wojciech Wislicki\\
{\em A. Soltan Institute for Nuclear Studies, Warszawa, Poland}\\
Pawel Moskal\\
{\em Jagiellonian University, Institute of Physics, Krakow, Poland}
}
\maketitle
\baselineskip=11.6pt
\begin{abstract}
The KLOE-2 detector is a powerful tool to study the temporal evolution
of quantum entangled pairs of kaons. The accuracy of such studies may
in principle be limited by the interaction of neutral kaons with
thermal photons present inside the detector. Therefore, it is crucial
to estimate the probability of this effect and its influence on the
interference patterns. In this paper we introduce the phenomenology
of the interaction of photons with neutral kaons and present and
discuss the obtained quantitative results.
\end{abstract}
\baselineskip=14pt
\section{The interaction frequency between thermal photons \\and neutral kaons}

Interaction between $K^0$ meson and thermal photons may remain undetected inside the KLOE-2
detector and constitute the background process where quantum coherence is destroyed.

To estimate the probability of this
interaction, we assume in the calculations that photons are in a room
temperature and $K^0$ is moving with respect to the laboratory frame
with the energy obtained in the $\phi\to K^0 \bar{K^0}$ decay.

\indent Meson $K^0$ is electrically neutral but it has inner
electromagnetic structure so it can interact with photons. We are
interested in interactions in which $K^0$ is observed in a final
state. The main process is the inverse Compton effect, that is
photon scattering on kaon in which photon's energy increases and
kaon's decreases. The total number of photons scattered on a kaon in
time unit is given by:
\begin{eqnarray}
\frac{dN}{dt}=\frac{1}{\gamma}\frac{dN^*}{dt^*}=\frac{c}{\gamma}\int
dn^* \sigma^{\ast}_{\gamma K}(k^{\ast})\label{eq:dN/dt},
\end{eqnarray}
where $\gamma$ is the Lorentz factor, $c$ the velocity of light and
$\sigma_{\gamma K}$ denotes the cross section for $\gamma K^0$
Compton scattering. Superscript ,,$*$'' indicates the rest frame of
$K^0$ meson. We denote by $dn$ number of photons in unit
of volume in the $dk$ energy interval, given by:
\begin{eqnarray}
dn=\frac{1}{4 \pi^{3}}\cdot \frac{k^{2}dk d\phi \sin\theta
d\theta}{e^{\frac{k}{k_{B}T}}-1}\label{eq:dn}.
\end{eqnarray}
The above formula was obtained assuming that the photon distribution
is given by the Planck's law of black-body radiation in the
temperature T. Here $\theta$ stands for the polar angle between the
incoming photon and the velocity of kaon, and $k_B$ is the Boltzmann constant.

\indent The cross section $\sigma_{\gamma K}$ for the $\gamma K^0$
Compton scattering can be obtained from the cross section for the
radiative scattering of the kaon in electromagnetic fields of
nuclei, known as the Primakoff effect, and is given by
\cite{Moinester}:
\begin{eqnarray}
\frac{d\sigma^{\ast}_{\gamma K}(k^{\ast},\theta^{\ast})}{
d\cos\theta^{\ast}}=\frac{2\pi\alpha_{f}}{m_K} \frac{k^{\ast
2}\Big(\alpha_{K}(1+\cos^{2}\theta^{\ast})+2\beta_{K}\cos\theta^{\ast}\Big)}
{\Big(1+\frac{k^{\ast}}{m_K}(1-\cos\theta^{\ast})\Big)^{3}}\label{eq:compton},
\end{eqnarray}
where $m_K$ is the $K^0$ mass and $\alpha_f$ the fine-structure constant. The $\alpha_K$ and $\beta_K$ stand for the electric and magnetic
polarizability of $K^0$. These quantities characterize
susceptibility of the $K^0$ to the electromagnetic field. Taking into account the Lorentz transformation of the photon
energy from the laboratory frame to the rest frame of $K^0$:
\begin{eqnarray}
k^*= \gamma k(1-\beta \cos\theta))\label{eq:kinw}
\end{eqnarray}
and the Lorentz invariance of $dn/k$, one gets the transformation of
the density of photons:
\begin{eqnarray}
dn^*=dn(1-\beta \cos \theta)\gamma,\label{eq:dninw}
\end{eqnarray}
where $\beta$ is the velocity of $K^0$ with respect to the
laboratory frame. Using consecutively equations (\ref{eq:dninw}),
(\ref{eq:dn}) and (\ref{eq:kinw}) and knowing that $\int_{0}^{2\pi}d\phi=2\pi$, the formula for the interaction
frequency (\ref{eq:dN/dt}) reads (where $u=\cos\theta$):
\begin{eqnarray}
\frac{dN}{dt}=\frac{c}{2 \pi^{2}}\int_{0}^{\infty}dk
\frac{k^{2}}{e^{\frac{k}{k_{B}T}}-1}\int_{-1}^{1}du \cdot(1-\beta
u)\cdot \sigma^{\ast}_{\gamma K}(\gamma k(1-\beta u)).\label{eq:czestosc}
\end{eqnarray}

\section{Units and values of parameters}

Numerical values of parameters $\alpha_{K}$ and $\beta_{K}$ used in equations in the last
paragraph are equal to $\alpha_{K}=2.4\cdot 10^{-49} \mbox{ m}^{3}$ and $\beta_{K}=10.3\cdot 10^{-49} \mbox{ m}^{3}$ \cite{Ebert}. Values for $\alpha_f, m_K, k_B$ and $c$ are taken from Particle Data Group \cite{pdg}. Temperature is assumed to be 300K. 

In natural units, the conversion $\mbox{eV}\to \mbox{m}$ should be done in
the following way: $\mbox{eV}=(197.33\cdot 10^{-9}\mbox{ m})^{-1}$, so the unit of (\ref{eq:czestosc}) is:
\begin{eqnarray}
\Big[\frac{dN}{dt}\Big]=\mbox{m}^4\cdot \mbox{eV}^4\cdot\frac{1}{\mbox{s}}=6.595\cdot 10^{26}\frac{1}{\mbox{s}}.
\label{eq:jednostka}
\end{eqnarray}

In the case of the $\phi\to K^0 \bar{K^0}$ decay the kinetic energy
of kaons in the laboratory frame is equal to ca. $E=12\mbox{ MeV}$, corresponding
to:\\

\begin{tabular}{c c c}
\centering
$\gamma=\frac{E+m_K}{m_K}=1.02412$, & & $\beta=\sqrt{1-\frac{1}{\gamma^{2}}}=0.21573$.\\
\end{tabular}

\section{Calculation of the cross section for inverse Compton scattering of $\gamma$ on $K^0$}

The total cross section $\sigma_{\gamma K}$ may be obtained by
integrating (\ref{eq:compton}) over the $\cos\theta^{\ast}$. In
order to simplify the calculations we will introduce the notation $\cos\theta^{\ast}=x$ and $u=-m-k^{\ast}+k^{\ast}x$:
\begin{eqnarray}
\sigma^{\ast}_{\gamma
K}(k^{\ast},u)&=&2\pi\alpha_{f}m^{2}\Big(\alpha_{K}\int\frac {-k^{\ast
2}-(m+k^{\ast})^{2}-u^{2}-2u(m+k^{\ast})}{k^{\ast}u^{3}}du+ \nonumber \\ 
&+&2\beta_{K}\int\frac{-m-k^{\ast}-u}{u^{3}}du \Big) \label{eq:sigma}.
\end{eqnarray}

After calculating $\sigma^{\ast}_{\gamma K}(k^{\ast},u)$ and replacing $u=-m-k^{\ast}+k^{\ast} x$, we integrate it over $x$ in the interval $[-1,1]$. As a result we get:
\begin{eqnarray}
\sigma^{\ast}_{\gamma K}(k^{\ast})&=&\sigma^{\ast}_{\gamma K}(k^{\ast},x\mid_{-1}^{1})=\frac{2\pi\alpha_{f}}{k^{\ast}(m+2k^{\ast})^2}\Big(2k^{\ast}(2(\alpha_{K}+\beta_{K})k^{\ast 3}+\nonumber\\
&-&3\alpha_{K}m^{2}k^{\ast}-\alpha_{K}m^{3})+\alpha_{K}m^{2}(2k^{\ast}+m)^{2}(\ln
\frac{m+2k^{\ast}}{m})\Big) \label{eq:sigma4}.
\end{eqnarray}

\section{Calculation of interaction frequency}
\begin{figure}[!h]
\centering
\includegraphics[width=3.4in]{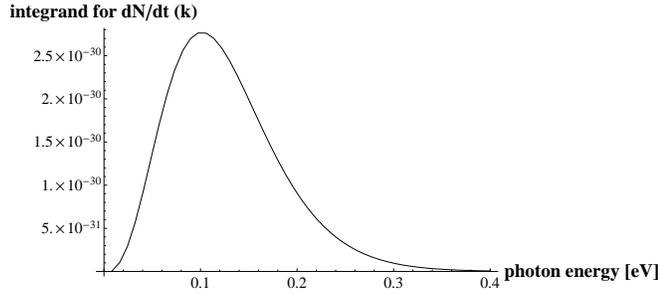} 
\caption{\it Integrand for $\frac{dN}{dt}(k)$}\label{integrand}
\end{figure}

Now we put equation (\ref{eq:sigma4}) from the previous section into formula for $\frac{dN}{dt}$ (\ref{eq:czestosc}). The integrand for $\frac{dN}{dt}$, multiplied by the unit conversion constant (\ref{eq:jednostka}), is shown in Figure \ref{integrand}.

Finally integrating numerically $\frac{dN}{dt}$ over $k$ we obtain:
\begin{eqnarray*}
\frac{dN}{dt}=3.7\cdot 10^{-31}\frac{1}{s}
\end{eqnarray*}

\subsection{Numerical stability}

Integral calculated in this chapter is quite sensitive to the numerical
accuracy and have to be treated with caution. The graph below shows the value of the whole integral $\frac{dN}{dt}$ (\ref{eq:czestosc}) with respect to the numerical precision (number of significant digits). One can see from it, that when we reach sufficient precision, the result stabilizes.

\begin{figure}[!h]
\centering
\includegraphics[width=3.7in]{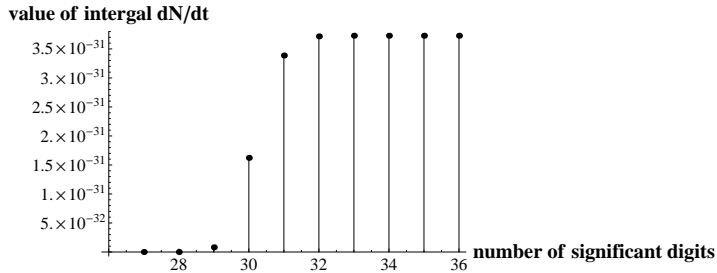}
\caption{\it Value of integral dN/dt}\label{integral}
\end{figure}

\subsection{Investigation on systematic errors}

Although the interaction probability is small, the systematic error on it was estimated. Obvious sources of systematics are uncertainty of $\alpha_K$ and $\beta_K$ and  variation of temperature. 

\begin{figure}[!h]
\centering
{\includegraphics[width=2.3in]{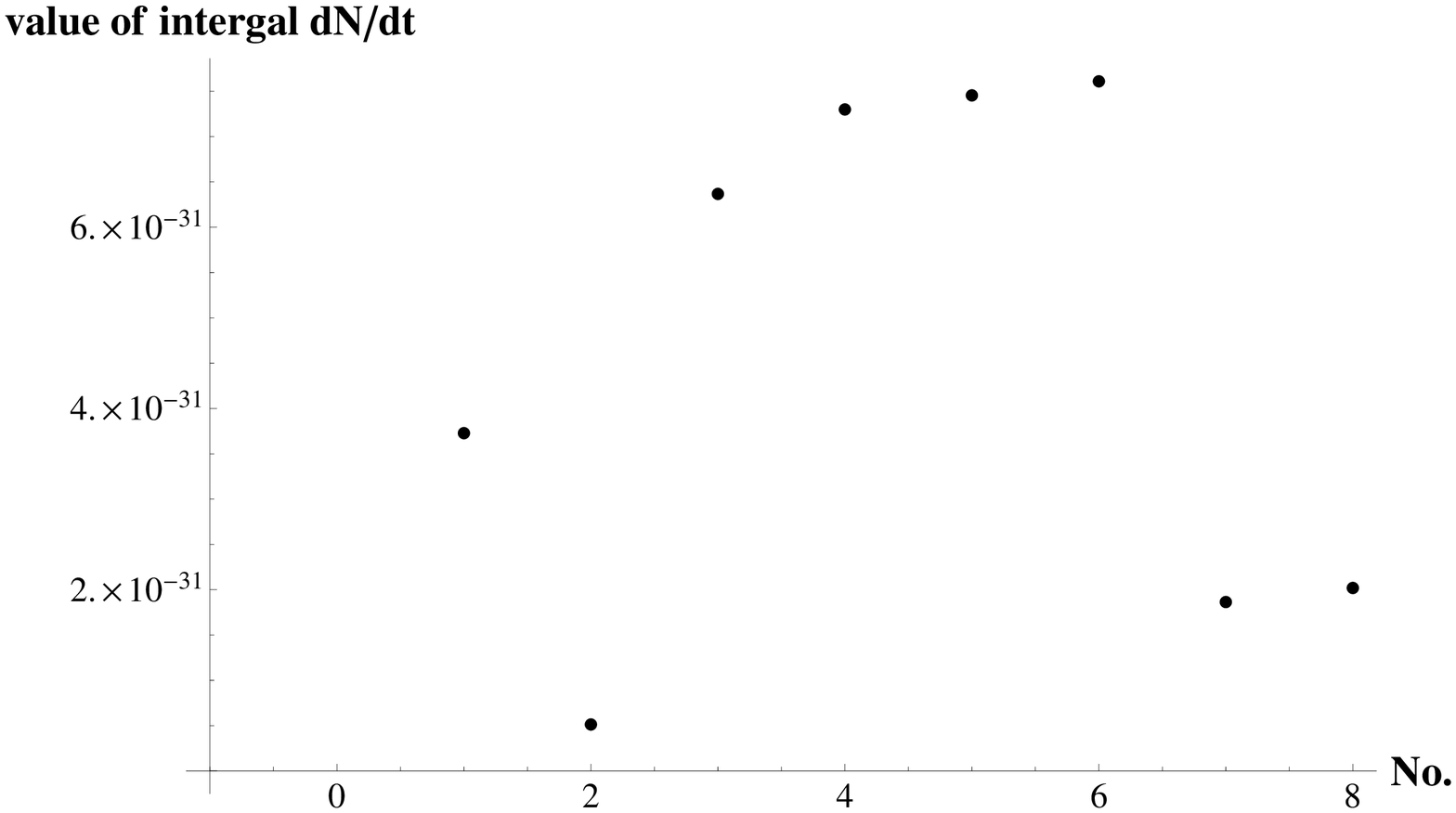}}
{\includegraphics[width=2.3in]{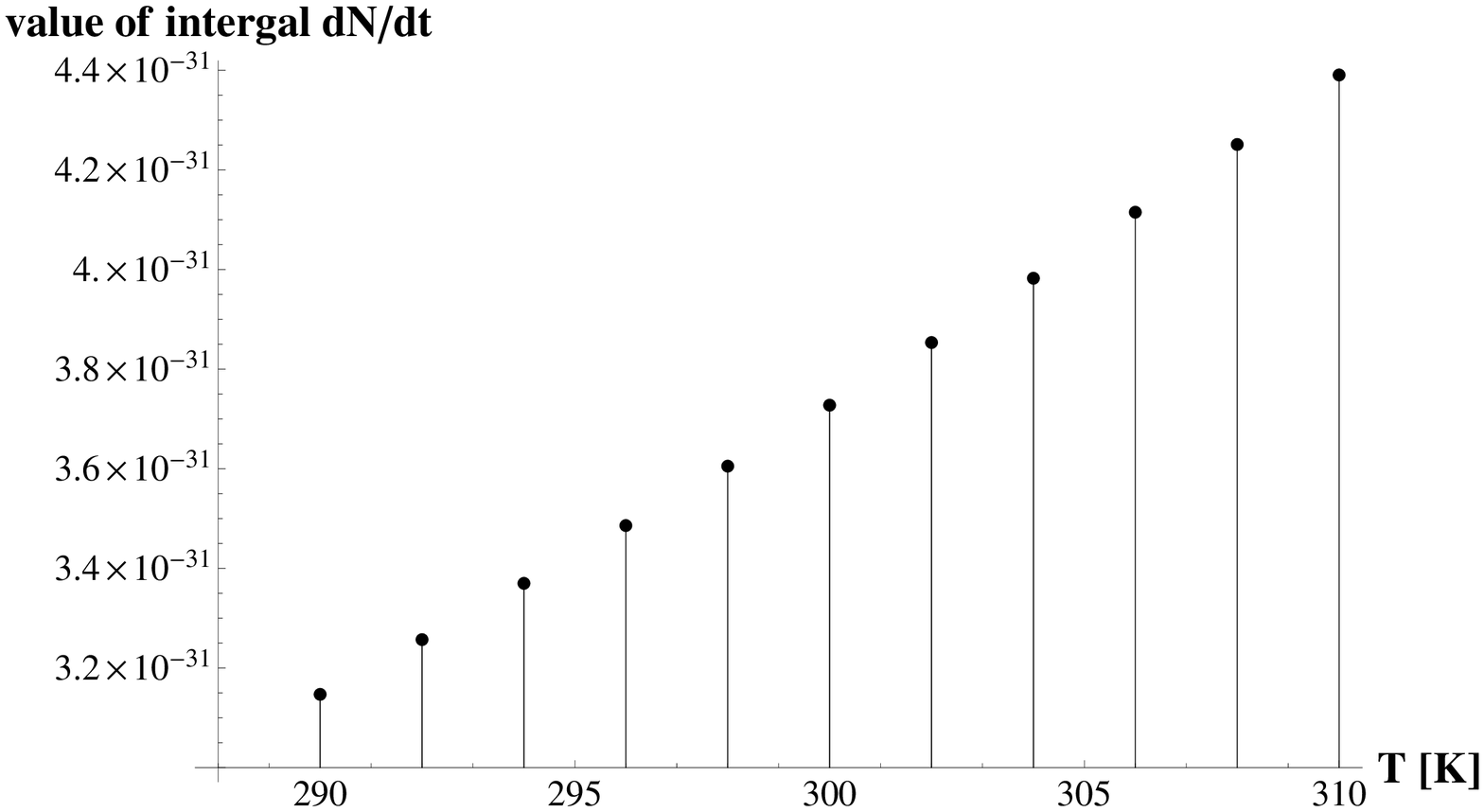}}
\caption{\it a) Values of integral for different $\alpha_K$ and $\beta_K$ parameters. b) Values of integral as a function of temperature.}
\label{error}
\end{figure}

The first one was estimated using values of $\alpha_K$ and $\beta_K$, derived using different methods in papers \cite{Wilcox} and \cite{Ivanov}. The result obtained in this paragraph was calculated using kaon polarizabilities taken from the paper \cite{Ebert}. Points on the graph \ref{error}a correspond to the different combinations of $\alpha_K$ and $\beta_K$. Figure \ref{error}b illustrates how the result changes due to the room temperature variations in the range of 20K around value of 300K.

Depending on the assumed values of $\alpha_K$, $\beta_K$ and temperature, the $\frac{dN}{dt}$ varies from about $5\cdot 10^{-32}$ to $7.5\cdot 10^{-31}$ so by more than one order of magnitude. However, as we will see in the next paragraph, this difference is not significant for the parameters of decoherence of kaon pairs at KLOE-2 detector.

\section{Physical interpretation}

Kaon is moving with velocity equal to ca. $v=0.6\cdot 10^8 \frac{\mbox{m}}{\mbox{s}}$ with
respect to the laboratory frame. From the place of its creation to
the calorimeter it moves through about $2.5$m so it needs for it about $4.2\cdot 10^{-8}$s. Because the
frequency of the Compton interaction is $3.7\cdot10^{-31} \frac{1}{\mbox{s}}$ so
probability of the interaction is:
$$P=1.5\cdot10^{-38}$$
This background stays small with respect to the statistical uncertainty of
decoherence parameters expected in KLOE-2 \cite{amelino}.

\section{Acknowledgements}

We acknowledge support by Polish Ministry of Science and Higher Education through the Grant No. 0469/B/H03/2009/37.

\end{document}